\documentclass{article}
\usepackage{akFPSpro}

\def\be{\beta}
\def\ga{\gamma}
\def\de{\delta}

\def\th{\theta}

\def\la{\lambda}

\def\ph{\phi}

\def\ch{\chi}

\def\om{\omega}
\def\Ga{\Gamma}
\def\De{\Delta}

\def\La{\Lambda}

\def\Ps{\Psi}
\def\Om{\Omega}

\def\cA{{\cal A}}

\def\fr#1#2{{{#1} \over {#2}}}
\def\frac#1#2{\textstyle{{{#1} \over {#2}}}}

\def\prt{\partial}

\def\ket#1{|{#1}\rangle}
\def\bra#1{\langle{#1}|}

\def\half{{\textstyle{1\over 2}}}
\def\lsim{\mathrel{\rlap{\lower4pt\hbox{\hskip1pt$\sim$}}
    \raise1pt\hbox{$<$}}}
\def\gsim{\mathrel{\rlap{\lower4pt\hbox{\hskip1pt$\sim$}}
    \raise1pt\hbox{$>$}}}

\def\Z{\hat Z}

\def\Re{\hbox{Re}\,}
\def\Im{\hbox{Im}\,}

\def\ol#1{\overline{#1}}
\newcommand{\beq}{\begin{equation}}
\newcommand{\eeq}{\end{equation}}
\newcommand{\bea}{\begin{eqnarray}}
\newcommand{\eea}{\end{eqnarray}}
\newcommand{\rf}[1]{(\ref{#1})}

\begin{document}

\title{CPT AND LORENTZ VIOLATION IN \\
KAONS AND OTHER SYSTEMS 
  }
\author{
V.\ Alan Kosteleck\'y \\
{\em Physics Department, Indiana University, 
         Bloomington, IN 47405, U.S.A.}
  }
\maketitle
\baselineskip=11.6pt
\begin{abstract}
This talk provides a status report on CPT violation 
in neutral-meson oscillations,
focusing on implications of
the CPT- and Lorentz-violating standard-model extension.
\end{abstract}

\baselineskip=14pt

\section{Introduction}
\label{intro}

The standard model of particle physics is invariant under
CPT and Lorentz symmetry.
However,
small observable violations might emerge 
from a more fundamental theory\cite{cpt98}.
Sensitive CPT tests for these effects can be performed
by taking advantage of the finely balanced natural interferometers
provided by the neutral-meson systems.

It is possible to parametrize
any indirect CPT violation in a neutral-meson oscillation
with a complex quantity\cite{lw}.
Here, this quantity is denoted by $\xi_P$,
where $P$ represents one of the neutral mesons $K$, $D$, $B_d$, $B_s$.
Assuming that $\xi_K$ is constant,
experiments on kaons\cite{e773}
have determined that its real and imaginary parts 
are no greater than about $10^{-4}$.

In conventional quantum field theory,
$\xi_K$ cannot be constant.
The point is that the CPT theorem shows that $\xi_K$ must be zero
unless Lorentz symmetry is broken,
while using an explicit and general 
Lorentz-violating standard-model extension\cite{ck}
to calculate $\xi_K$ shows 
that it varies with the meson 4-momentum\cite{ak,ak2}.
This implies qualitatively new experimental tests of CPT
are possible,
one of which has recently been performed in the $K$ system\cite{k99}.
Various CPT tests have also been proposed\cite{kp,msas}
using the heavy mesons $D$, $B_d$, $B_s$.
Recent experiments\cite{bexpt}
have obtained bounds on 
$\Re\xi_{B_d}$ of order 1 
and on $\Im\xi_{B_d}$ of order $10^{-1}$
under the assumption of constant $\xi_{B_d}$.

This talk reviews the present theoretical situation
for CPT violation in neutral-meson systems
in the context of the CPT- and Lorentz-violating
standard-model extension.
Some experimentally accessible asymmetries are presented
for both uncorrelated and correlated neutral-meson systems,
and the implications of the variation of $\xi_P$ with 
meson 4-momentum are discussed.

\section{Basics}
\label{basics}

Any linear combination of the Schr\"odinger wave functions 
for a meson $P^0$ and its antimeson $\overline{P^0}$
can be represented as a two-component object $\Ps(t)$.
The time evolution of an arbitrary neutral-meson state 
is then controlled by a 2$\times$2 effective hamiltonian $\La$ 
according to
$i\prt_t \Ps = \La \Ps$.
The eigenstates $\ket{P_a}$ and $\ket{P_b}$
of $\La$ are physical states,
in analogy with the normal modes 
of a classical oscillator\cite{osc}.
They evolve as
$\ket{P_a(t)}=\exp (-i\la_at) \ket{P_a}$,
$\ket{P_b(t)}=\exp (-i\la_bt) \ket{P_b}$,
where the complex parameters 
$\la_a \equiv m_a - \half i \ga_a$,
$\la_b \equiv m_b - \half i \ga_b$
are the eigenvalues of $\La$,
with $m_a$, $m_b$ the physical masses
and $\ga_a$, $\ga_b$ the decay rates.
It is convenient to introduce the definitions
$\la \equiv \la_a + \la_b = m - \half i \ga$,
$\De \la \equiv \la_a - \la_b = - \De m - \half i \De \ga$,
where $m = m_a + m_b$, $\De m = m_b - m_a$,
$\ga = \ga_a + \ga_b$, $\De \ga = \ga_a - \ga_b$.

Without loss of generality,
the effective hamiltonian $\La$ can be
adopted as\cite{ak2,ll}
\beq
\La = 
\half \De\la
\left( \begin{array}{lr}
U + \xi 
& 
\quad VW^{-1} 
\\ & \\
VW \quad 
& 
U - \xi 
\end{array}
\right),
\label{uvwx}
\eeq
where the parameters $UVW\xi$ are complex.
The prefactor $\De\la/2$ 
ensures these parameters are dimensionless
and eliminates some factors of 2 in subsequent equations.
Imposing the trace as tr$~\La = \la$ 
and the determinant as $\det \La = \la_a \la_b$
shows that
$U \equiv \la/\De\la$
and $V \equiv \sqrt{1 - \xi^2}$.

The independent complex parameters 
$W = w \exp (i\om)$,
$\xi = \Re\xi + i \Im \xi$
in Eq.\ \rf{uvwx} 
have four real components.
However,
one is physically unobservable:
the argument $\om$ changes under a phase redefinition 
of the $P^0$ wave function.
The three others are physical.
The parameter $w$ determines the amount of T violation,
with T preserved if and only if $w = 1$.
The two real numbers
$\Re\xi$, $\Im\xi$
determine the amount of CPT violation,
with CPT preserved if and only if both are zero.
Note that in the standard notation 
specific to the $K$ system,
in which the complex parameter for CPT violation is
often denoted $\de_K$,
imposing small CP violation and making
a suitable choice of phase convention yields
the identification $\xi_K\approx 2\de_K$.

The three CP-violation parameters $w$, $\Re\xi$, $\Im\xi$
in this $w\xi$ formalism
are dimensionless,
can be used for arbitrary size CPT and T violation,
and are independent of phase conventions.
They are also independent of any specific model
because they are phenomenological.
However,
it is unjustified \it a priori \rm
to suppose that they must be constant numbers.
In fact,
the assumption often found in the literature
that $\xi$ is constant and nonzero
is an additional strong requirement,
which the CPT theorem shows is inconsistent 
with the basic axioms of Lorentz-invariant quantum field theory.
If instead Lorentz violations are allowed,
then in quantum field theory $\xi$ cannot be constant 
and is found to vary with the meson 4-momentum.
This result is outlined in the next part of the talk.

\section{Theory for CPT Violation}
\label{theory}

Lorentz-invariant quantum field theories
are CPT-symmetric by virtue of the CPT theorem.
In describing CPT violation at the level of quantum field theory,
it is thus interesting to
study the consequences of small Lorentz violations.
A general CPT- and Lorentz-violating standard-model extension 
exists\cite{ck} and could arise,
for instance,
as the low-energy limit of an underlying Planck-scale theory\cite{kps}.
In addition to the neutral-meson oscillations discussed here,
signals in a variety of other types of experiment
are predicted by the standard-model extension.
These include,
for example,
tests of quantum electrodynamics with trapped particles\cite{bkr},
measurements of muon properties\cite{bkl},
hydrogen and antihydrogen spectroscopy\cite{bkr2},
clock-comparison experiments\cite{kla},
studies of the behavior of a spin-polarized torsion pendulum\cite{bk},
measurements of cosmological birefringence\cite{cfj},
and observations of the baryon asymmetry\cite{bckp}.
However,
none of these experiments involve flavor-changing effects,
and as a result it can be shown that they leave unconstrained
the sector of the standard-model extension
relevant to experiments with neutral-meson oscillations\cite{ak}.

The dominant CPT-violating contributions to 
the effective hamiltonian $\La$
can be calculated as expectation values of interaction terms
in the standard-model extension.
It can be shown that
the difference 
$\De\La =\La_{11} - \La_{22}$
of the diagonal terms of $\La$
is given by\cite{ak} 
\beq
\De\La \approx \be^\mu \De a_\mu
\quad ,
\label{dem}
\eeq
where $\be^\mu = \ga (1, \vec \be )$ is the four-velocity
of the $P$ meson in the laboratory frame
and the coefficients $\De a_\mu$
are combinations of coefficients appearing in 
the lagrangian for the standard-model extension.

There are four independent components in $\De a_\mu$,
which implies that a complete characterization of CPT violation 
requires four independent CPT measurements 
in each $P$-meson system.
Moreover,
the 4-velocity and consequent 4-momentum dependence in Eq.\ \rf{dem} 
shows explicitly that CPT violation cannot be described 
with a constant complex parameter in quantum field theory.
Since the effects from CPT violation will typically vary 
with the momentum magnitude and orientation of the $P$ mesons,
the experimental reach depends on 
the meson momentum spectrum and angular distribution\cite{ak,ak2}.
Among other consequences,
this implies that experiments previously regarded as 
equivalent may in fact have different CPT reaches.

Another feature of experimental importance is
the variation of some CPT observables with sidereal time\cite{ak,ak2},
resulting from the rotation of the Earth 
relative to the constant vector $\De\vec a$.
To exhibit directly the sidereal-time dependence,
it is necessary to convert the result \rf{dem} for $\De\La$ 
from the laboratory frame rotating with the Earth
to a nonrotating frame.
It is convenient to adopt a nonrotating frame
compatible with celestial equatorial coordinates.
Let the coefficient $\vec a$ for Lorentz violation in a $P$-meson system
have nonrotating-frame components $(a^X, a^Y, a^Z)$.
Take the unit vector $\Z$ to be aligned 
along the Earth's rotation axis,
and let $\vec\be = \be (\sin\th\cos\ph, \sin\th\sin\ph, \cos\th)$
be the laboratory-frame 3-velocity of a $P$ meson,
where the angles $\th$, $\ph$ are defined with respect to
the laboratory-frame $\hat z$ axis.
Define the momentum magnitude $p \equiv |\vec p| =\be m_P \ga(p)$,
where $\ga(p) = \sqrt{1 + p^2/m_P^2}$ as usual.
Then,
it can be shown that in any $P$ system 
and for arbitrary size CPT violation
the complex CPT parameter $\xi$ 
is\cite{ak2}
\bea
\xi &\equiv &
\xi(\hat t, \vec p) \equiv \xi(\hat t, p, \th, \ph) 
\nonumber\\
&=& 
\fr 
{\ga( p)}
{\De \la} 
\bigl\{
\De a_0 
+ \be \De a_Z 
(\cos\th\cos\ch - \sin\th \cos\ph\sin\ch)
\nonumber\\
&&
\qquad
+\be \bigl[
\De a_Y (\cos\th\sin\ch 
+ \sin\th\cos\ph\cos\ch )
-\De a_X \sin\th\sin\ph 
\bigr] \sin\Om \hat t
\nonumber\\
&&
\qquad
+\be \bigl[
\De a_X (\cos\th\sin\ch 
+ \sin\th\cos\ph\cos\ch )
+\De a_Y \sin\th\sin\ph 
\bigr] \cos\Om \hat t
\bigr\} ,
\nonumber\\
\label{xipt}
\eea
where $\hat t$ is the sidereal time.
In the next part of the talk,
some implications of the expression \rf{xipt} for experiment
are presented.

\section{Experiment}
\label{expt}

Consider for simplicity the case of semileptonic decays
into a final state $f$ or its conjugate state $\overline{f}$,
neglecting any violations of the 
$\De Q = \De S$,
$\De Q = \De C$,
or $\De Q = \De B$ rules.
The basic transition amplitudes can be taken as 
$\bra{f}T\ket{P^0} = F$,
$\bra {\overline f}T\ket {\overline{P^0}} = \overline F$,
$\bra{f}T\ket{\overline{P^0}} 
=\bra {\overline f}T\ket{P^0} = 0$.
Time-dependent decay amplitudes and probabilities can then be calculated
as usual.
In addition to the proper-time dependence,
there is now also sidereal time and momentum dependence 
from $\xi(\hat t, \vec p)$.
Note that $\xi$ is independent of $t$
to an excellent approximation 
because the meson decays are rapid on the scale of sidereal time.

As a simple example for uncorrelated mesons,
consider the case where $F^* = \ol F$,
i.e., negligible direct CPT violation.
In terms of decay probabilities,
a CPT-sensitive asymmetry is
\bea
\cA^{\rm CPT}(t,\hat t,\vec p) &\equiv& 
\fr{
\overline{P}_{\overline{f}}(t,\hat t,\vec p) - P_f(t,\hat t,\vec p) 
}{
\overline{P}_{\overline{f}}(t,\hat t,\vec p) + P_f(t,\hat t,\vec p) 
}
\nonumber\\
&&
=\fr{
 2 \Re \xi \sinh\De\ga t/2
+ 2 \Im \xi \sin \De m t
}{
(1 + |\xi|^2) \cosh\De\ga t/2
+(1 - |\xi|^2) \cos\De m t
},
\label{corrasymm}
\eea
which depends implicitly on 
$\hat t$, $\vec p$ through the dependence on $\xi (\hat t, \vec p)$.

Careful averaging over one of more of the variables
$t$, $\hat t$, $p$, $\th$, $\ph$
either before or after constructing the asymmetry \rf{corrasymm}
yields independent bounds on the four coefficients $\De a_\mu$.
In particular,
inspection of Eq.\ \rf{xipt} reveals that binning 
in $\hat t$ gives information on $\De a_X$ and $\De a_Y$,
while binning in $\th$ separates the spatial and timelike components 
of $\De a_\mu$.
An example that has already given two independent CPT bounds
of about $10^{-20}$ GeV each on different combinations 
of the coefficients $\De a_\mu$
in the $K$ system\cite{ak,k99} is provided
by the special case of mesons highly collimated in the laboratory frame,
for which the 3-velocity can be written $\vec\be = (0,0,\be )$
and $\xi$ simplifies.
Binning in $\hat t$ provides sensitivity to 
the equatorial components $\De a_X$, $\De a_Y$,
while averaging over $\hat t$ eliminates them altogether.
More generally, 
note that the variation with sidereal time can provide clean CPT bounds
even using observables that mix T and CPT effects\cite{ak2},
such as the standard rate asymmetry $\de_l$ for $K_L$ semileptonic decays.

As another example,
consider the case of the decay into $f\overline{f}$
of a correlated meson pair created by quarkonium decay.
The probability for the double decay
is a function of the sidereal time $\hat t$
and of the proper decay times $t_1$, $t_2$ 
and momenta $\vec p_1$, $\vec p_2$
of the two mesons.
Note that according to Eq.\ \rf{xipt}
the CPT-violating parameters $\xi_1$ and $\xi_2$
for each meson typically differ.
Experimentally,
the time sum $t = t_1 + t_2$ is unobservable
and so the relevant probability $\Ga_{f\ol f}$
is found by integrating over $t$.
An asymmetry $\cA^{\rm CPT}_{f\ol f}$ sensitive to 
the sum $\xi_1 + \xi_2$ 
and the difference $\De t = t_1 - t_2$
can then be defined as 
\bea
\cA^{\rm CPT}_{f\ol f}(\De t, \hat t, \vec p_1, \vec p_2)
&=&
\fr{
\Ga _{f\ol f}(\De t, \hat t, \vec p_1, \vec p_2) 
- \Ga _{f\ol f}(-\De t, \hat t, \vec p_1, \vec p_2)
}{
\Ga _{f\ol f}(\De t, \hat t, \vec p_1, \vec p_2)
+ \Ga _{f\ol f}(-\De t, \hat t, \vec p_1, \vec p_2)
}
\nonumber\\
&=&
\fr{
-\Re (\xi_1 + \xi_2)
\sinh \half \De\ga \De t
-\Im(\xi_1 + \xi_2)
\sin\De m\De t
}{
\cosh \half \De\ga \De t
+
\cos \De m\De t
},
\nonumber\\
\label{sumxiasymm}
\eea
where the sidereal-time and momenta dependences are
implicit in $\xi_1$, $\xi_2$.
As before,
different experiments using this asymmetry may have different CPT reach.
If the quarkonium is produced at rest in a symmetric collider,
for instance,
then the sum $\xi_1 + \xi_2 = 2 \ga (p)\De a_0/\De\la$
is independent of $\De \vec a$,
so a direct fit to the variation with $\De t$
provides a bound on $\De a_0$.
In contrast,
quarkonium production in an asymmetric collider
implies $\xi_1 + \xi_2$ is sensitive 
to all four components of $\De a_\mu$,
and so suitable binning permits the extraction of four independent 
CPT bounds.
These experiments are feasible,
for example,
at the existing asymmetric $B_d$ factories BaBar and BELLE.

\section*{Acknowledgments}

This work was supported in part 
by the United States Department of Energy 
under grant number DE-FG02-91ER40661.

\end{document}